\documentclass{iopart}

\usepackage{graphicx}
\usepackage{mathrsfs}
\usepackage{hyperref}
\usepackage{geometry}
\usepackage{bm}
\newcommand{\text}[1]{\mathrm{#1}}

\begin{document}

\title{Reversible to Irreversible Transitions for ac Driven Skyrmions on Periodic Substrates
}
\author{J. C. Bellizotti Souza$^{1}$,
            N. P. Vizarim$^{2}$, 
            C. J. O. Reichhardt$^3$, 
            C. Reichhardt$^3$
            and P. A. Venegas$^4$}
            
\ead{jc.souza@unesp.br}
\address{$^1$ POSMAT - Programa de P\'os-Gradua\c{c}\~ao em Ci\^encia e Tecnologia de Materiais, S\~ao Paulo State University (UNESP), School of Sciences, Bauru 17033-360, SP, Brazil}

\address{$^2$ Departament of Electronics and Telecommunications Engineering, S\~ao Paulo State University (UNESP), School of Engineering, S\~ao J\~oao da Boa Vista 13876-750, SP, Brazil}

\address{$^3$ Theoretical Division and Center for Nonlinear Studies, Los Alamos National Laboratory, Los Alamos, New Mexico 87545, USA}

\address{$^4$ Department of Physics, S\~ao Paulo State University (UNESP), School of Sciences, Bauru 17033-360, SP, Brazil}

\date{\today}

\begin{abstract}

Using atomistic simulations, we investigate the dynamical behavior of 
magnetic skyrmions in dimer and trimer molecular crystal arrangements,
as well as bipartite lattices at 3/2 and 5/2 fillings, under ac driving
over a square array of anisotropy defects.
For low ac amplitudes,
at all fillings we find reversible motion where the skyrmions
return to their original positions at the end of each ac drive cycle
and the diffusion is zero.
We also identify two distinct irreversible regimes.
The first is a translating regime in which the skyrmions form
channels of flow in opposing directions and translate by one substrate
lattice constant per ac drive cycle.
The translating state appears in the dimer and trimer states,
and produces
pronounced peaks in the diffusivity in the direction perpendicular
to the external drive.
For larger ac amplitudes, we find chaotic irreversible motion
in which the skyrmions can randomly exchange
places with each other over time,
producing
long-time diffusive behavior both parallel and perpendicular to the
ac driving direction.
\end{abstract}

\maketitle

\section{Introduction}
A large variety of dynamic phenomena can emerge for
many body systems under dc driving over
random or periodic substrates \cite{reichhardt_depinning_2017}.
Examples of such systems include colloidal particles
\cite{pertsinidis_statics_2008,tierno_depinning_2012},
type-II superconducting vortices
\cite{bhattacharya_dynamics_1993,shaw_critical_2012},
active matter \cite{bechinger_active_2016},
sliding friction \cite{vanossi_colloquium_2013},
and geological systems \cite{carlson_properties_1989}.
The flows may be
strongly disordered
or chaotic in the plastic flow regimes
but ordered in elastic flow regimes \cite{reichhardt_depinning_2017}.
If ac rather than dc driving is applied, even in the absence of a substrate
it is possible for many body systems to undergo a transition from
reversible behavior,
where the particles return to the same positions
after each ac drive cycle,
to irreversible behavior, where the particles never return to their
original positions.
The irreversible flows can produce
long time diffusive behavior even in the absence of
thermal fluctuations.
The transition from reversible to irreversible behavior was
first studied in detail for periodically sheared
colloidal particles \cite{pine_chaos_2005}, where the net
displacement of the particles was measured after each shear cycle.
In the reversible regime, the colloids returned
to their initial configuration at the end of each cycle.
In the irreversible regime,
the configuration of the colloids changed from one cycle to the next,
with the colloids exhibiting chaotic dynamics and
undergoing long-term diffusive behavior after many cycles.
Pine {\it et al.}~\cite{pine_chaos_2005} found that for fixed colloidal density,
there was a critical ac drive amplitude above which irreversible
behavior appeared, while if the ac amplitude was fixed, there
was a critical density above which the motion became irreversible.

In further investigations of the periodically sheared
colloidal system,
Corte {\it et al.}~\cite{corte_random_2008} found that
while the initial motion is always irreversible,
the particles eventually organize into a steady state that may be either
reversible or irreversible, depending on the distance traveled during
the shearing process.
Notably, the number of cycles required to reach this
steady state diverges as a power law at a critical point,
indicating that the reversible-irreversible transition
is a nonequilibrium phase transition.
In the colloidal system, the behavior becomes reversible when the
particles are able to organize into a state where 
particle-particle collisions never occur.
Additional investigations
of reversible to irreversible (R-IR) transitions have since been performed
in other systems with much stronger particle-particle interactions,
where interactions between the particles remain important even in the
reversible state.
These include 
granular matter \cite{schreck_particle-scale_2013, milz_connecting_2013},
dislocations \cite{zhou_random_2014,ni_yield_2019}, amorphous solids 
\cite{regev_onset_2013,regev_reversibility_2015,priezjev_reversible_2016,leishangthem_yielding_2017}, 
polycrystalline solids \cite{jana_irreversibility_2017}, 
type-II superconducting vortices \cite{mangan_reversible_2008,okuma_transition_2011,dobroka_memory_2017} 
and, more recently, magnetic skyrmions \cite{brown_reversible_2019}. 

Previous work on R-IR transitions for
superconducting vortices and
magnetic skyrmions was performed under coupling
to a randomly disordered substrate.
Far less is known about
the nature of R-IR transitions for systems on 
periodic substrates \cite{Reichhardt23}, where
commensuration effects can arise
when the ratio of the number of particles to the number of substrate
minima is an integer or a rational fraction
\cite{Harada96,reichhardt_commensurate_1998,Bohlein12,reichhardt_depinning_2017}.
For commensurate conditions,
the system generally adopts an ordered structure.
This implies that under periodic shearing, different effects could arise
compared to what is found for systems with random disorder.
In this work we study R-IR transitions
for magnetic skyrmions subjected to ac driving over periodic substrates for
integer and rational fillings.
Compared to
the previous work on R-IR transitions in skyrmion systems
\cite{brown_reversible_2019}, the two key differences in the present work
are that we consider a periodic substrate,
and that instead of using a
point particle or Thiele equation approach where the
skyrmions have no internal degrees of freedom, we perform
atomistic simulations that permit
the skyrmions to change both shape and size.


Magnetic skyrmions are particle-like topologically
protected magnetic textures \cite{nagaosa_topological_2013, je_direct_2020}
that exhibit many similarities to overdamped particles.
Both minimize their
repulsive interactions by forming a triangular array, can be set in motion
by the application of external driving,
and can interact with material defects in
several distinct ways \cite{olson_reichhardt_comparing_2014,reichhardt_depinning_2017,reichhardt_statics_2022}.
The key difference between skyrmions and
other overdamped particles
is the presence of a non-dissipative Magnus force
that causes the
skyrmions, in clean samples, to move at an angle known as the intrinsic
skyrmion Hall angle, $\theta^\text{int}_\text{sk}$, with respect to the external
driving force
\cite{nagaosa_topological_2013, litzius_skyrmion_2017, iwasaki_universal_2013, jiang_direct_2017, lin_driven_2013, lin_particle_2013}.
This Magnus force produces richer dynamical behavior,
with a greater number of possible dynamical phases,
for the skyrmions compared to other overdamped particles,
since the skyrmion motion is not along the drive direction,
but follows a finite angle.
The Magnus force can also significantly change the interaction of
the skyrmions with a substrate.
In the limit
of large Magnus forces, the skyrmion moves parallel
rather than perpendicular to equipotential lines of the substrate
potential
\cite{feilhauer_controlled_2020}.
For more moderate Magnus forces, the skyrmion picks up
a velocity component that is perpendicular to the substrate force
\cite{reichhardt_statics_2022}.
The rich dynamical behavior of skyrmions interacting with
substrates has been 
intensively investigated in recent years, spanning topics that include
periodic pinning \cite{reichhardt_quantized_2015, reichhardt_nonequilibrium_2018, feilhauer_controlled_2020, vizarim_directional_2021, vizarim_skyrmion_2020, reichhardt_commensuration_2022},
interface guided motion \cite{vizarim_guided_2021, zhang_edge-guided_2022},
ratchet effects \cite{reichhardt_magnus-induced_2015, souza_skyrmion_2021, chen_skyrmion_2019, gobel_skyrmion_2021, souza_controlled_2024},
temperature and magnetic field gradients \cite{yanes_skyrmion_2019, zhang_manipulation_2018, everschor_rotating_2012, kong_dynamics_2013},
granular films \cite{gong_current-driven_2020,del-valle_defect_2022},
parametric pumping \cite{yuan_wiggling_2019},
voltage-controlled perpendicular magnetic anisotropy (PMA)
\cite{zhang_magnetic_2015,zhao_ferromagnetic_2020},
skyrmion lattice compression \cite{zhang_structural_2022, bellizotti_souza_spontaneous_2023},
soliton motion along skyrmion chains \cite{vizarim_soliton_2022, souza_soliton_2023},
pulse current modulation \cite{du_steady_2024},
strain driven motion \cite{liu_strain-induced_2024}, and
motion through interaction with domain walls \cite{xing_skyrmion_2022}.
Skyrmions are bubble-like objects that can change shape and size, but
most previous studies of R-IR transitions
have focused on particles that have a fixed
size and shape.

Recently, Souza {\it et al.}~\cite{souza_skyrmion_2024} reported
the stabilization of skyrmion molecular crystals on triangular
substrates. The skyrmion molecular crystals exhibited
the same ordering as other molecular crystals found for
superconducting vortices
\cite{berdiyorov_vortex_2006,neal_competing_2007,reichhardt_vortex_2007},
colloidal particles \cite{reichhardt_novel_2002,brunner_phase_2002,agra_theory_2004,frey_melting_2005,thomas_spin_2007}, 
and Wigner crystals \cite{Reddy23,Li24}.
These arrangements of particles are known as ``molecular crystals'' 
because, at integer fillings, particles trapped in individual substrate
minima
organize into dimer, trimer,
and higher-order \textit{n}-mer states that can
exhibit an overall orientational order \cite{reichhardt_vortex_2007}.
The dynamics of the particles in these dimer, trimer, or \textit{n}-mer
states could lead to novel behavior at R-IR transitions.

Using atomistic simulations, we simulate the dynamical
behavior of different skyrmion molecular crystal arrangements,
including dimers, trimers, and bipartite lattices
with 3/2 and 5/2 fillings, under ac driving over
a square array of sites with modified anisotropy.
We measure the positions of the skyrmions after each
ac drive cycle.
In the irreversible regimes, the skyrmions change positions so that there
is a finite long-time diffusivity,
while in the reversible state, the diffusion is zero.
At low ac amplitudes, we observe
reversible skyrmion motion for all 
skyrmion arrangements, but we find that the ac amplitude
range over which the motion remains reversible
changes depending on the filling fraction.
For higher ac amplitudes, we find
peaks in the $y$-direction diffusivity, perpendicular to the driving
direction,
only for dimer and trimer fillings.
At these peaks,
the motion is irreversible since the diffusion is nonzero,
and a portion of the skyrmions translate by
a fixed upward or downward amount during each ac drive cycle
in the form of an edge current.
Even though
the motion is irreversible, the trajectories are ordered.
In this case, there
is no net directed motion of skyrmions
since half of the mobile skyrmions are moving in the $+y$ direction and
the other half are moving in the $-y$ direction,
so the system is not a ratchet.
We also find that by using other fillings or applying even larger
ac drive amplitudes, 
the skyrmions undergo chaotic irreversible
motion with a finite diffusivity that is much lower than the diffusivity
in the irreversible states with
directed motion.

\section{Methods}
We employ atomistic simulations,
which track the motion of individual atomic magnetic
moments \cite{evans_atomistic_2018}, to
model N{\' e}el skyrmions in an ultrathin
ferromagnetic sample.
The sample is of size 84~nm $\times$ 84~nm and has periodic
boundary conditions along the $x$ and $y$ directions.
The sample anisotropy
is given by
$K(x, y)=\frac{K_0}{4}\left[\cos\left(\frac{2\pi x}{a_0}\right)+\cos\left(\frac{2\pi y}{a_0}\right) + 2\right]$, where
$K_0$ is the anisotropy depth, $a_0=14$~nm is the substrate lattice constant,
and there are $N_m=36$ minima of the anisotropy. This
produces a substrate consisting of a periodic square array of defects.
We apply a magnetic field perpendicular to the sample
along $-z$ at zero temperature $T=0$~K.

\begin{figure}
    \centering
    \includegraphics[width=0.6\columnwidth]{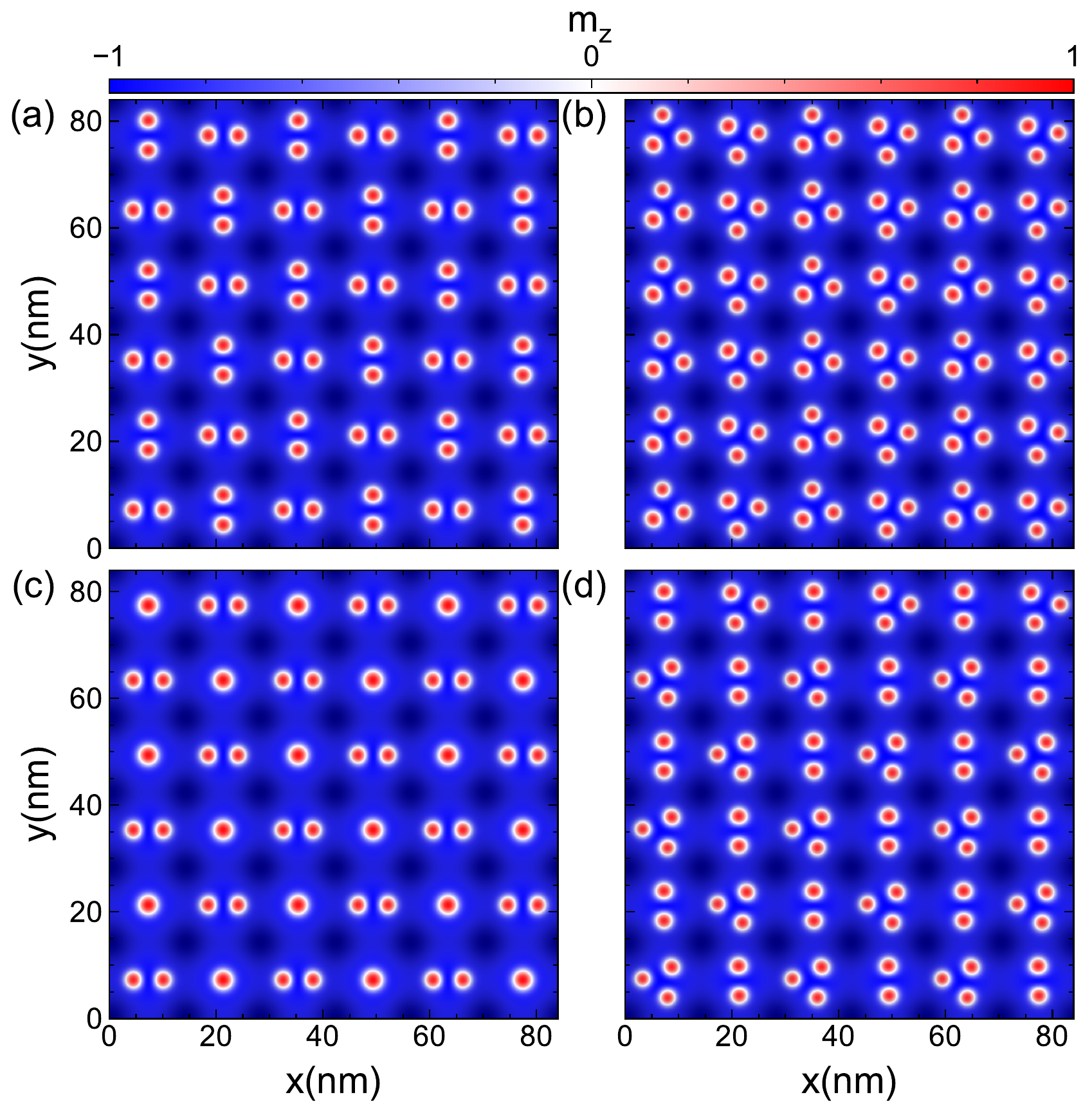}
    \caption{Images of the different initial skyrmion lattice arrangements
    used in this work. The PMA is plotted as a transparency
    overlay, with dark regions indicating high PMA.
    (a) Dimers forming an alternating pattern, or antiferromagnetic 
    pattern, at $N_\text{sk}/N_m=2$
    (b) Trimers in an alternating column pattern at $N_\text{sk}/N_m=3$.
    (c) Bipartite lattice containing monomers and dimers at
    $N_\text{sk}/N_m=3/2$.
    (d) Bipartite lattice containing dimers and trimers at
    $N_\text{sk}/N_m=5/2$.}
    \label{fig:1}
\end{figure}

The Hamiltonian governing the atomistic dynamics is given by
\cite{evans_atomistic_2018, iwasaki_universal_2013, iwasaki_current-induced_2013}:

\begin{eqnarray}\label{eq:1}
  \mathcal{H}=&-\sum_{i, j\in N}J_{ij}\mathbf{m}_i\cdot\mathbf{m}_j
                -\sum_{i, j\in N}\mathbf{D}_{ij}\cdot\left(\mathbf{m}_i\times\mathbf{m}_j\right)\\\nonumber
                &-\sum_i\mu\mathbf{H}\cdot\mathbf{m}_i
                -\sum_{i} K(x_i, y_i)\left(\mathbf{m}_i\cdot\hat{\mathbf{z}}\right)^2 \ . \\\nonumber
\end{eqnarray}
The underlying lattice is a square arrangement of magnetic moments
with lattice constant $a=0.5$~nm.
The first term on the right hand side is the exchange interaction
between the nearest neighbors contained in the set $N$,
with an exchange constant of $J_{ij}=J$ between magnetic moments
$i$ and $j$.
The second term is the interfacial Dzyaloshinskii–Moriya
interaction, where $\mathbf{D}_{ij}=D\mathbf{\hat{z}}\times\mathbf{\hat{r}}_{ij}$ is the Dzyaloshinskii–Moriya
vector between magnetic moments $i$ and $j$ and $\mathbf{\hat{r}}_{ij}$
is the
unit distance vector between sites $i$ and $j$.
The third term is the Zeeman interaction with an applied external magnetic
field $\mathbf{H}$.
Here $\mu=\hbar\gamma$ is the magnitude of the magnetic moment
and $\gamma=1.76\times10^{11}~$T$^{-1}$~s$^{-1}$ is the electron
gyromagnetic ratio. The last term represents the 
perpendicular magnetic anisotropy (PMA) of the sample, where
$x_i$ and $y_i$ are the spatial coordinates of the $i$th
magnetic moment.
Since we are considering ultrathin films, long-range dipolar interactions
are small enough that they can be neglected \cite{paul_role_2020}.

The time evolution for the individual
atomic magnetic moments is obtained using the LLG
equation \cite{seki_skyrmions_2016, gilbert_phenomenological_2004}:

\begin{equation}\label{eq:2}
    \frac{\partial\mathbf{m}_i}{\partial t}=-\gamma\mathbf{m}_i\times\mathbf{H}^\text{eff}_i
                             +\alpha\mathbf{m}_i\times\frac{\partial\mathbf{m}_i}{\partial t}
                             +\frac{pa^3}{2e}\left(\mathbf{j}\cdot\nabla\right)\mathbf{m}_i \ .
\end{equation}
Here $\gamma$ is the electron gyromagnetic ratio,
$\mathbf{H}^\text{eff}_i=-\frac{1}{\hbar\gamma}\frac{\partial \mathcal{H}}{\partial \mathbf{m}_i}$
is the effective magnetic field including all interactions from
the Hamiltonian, $\alpha$ is the phenomenological damping
introduced by Gilbert, and the last term is the adiabatic spin-transfer-torque
(STT), where $p$ is the spin polarization, $e$ the electron
charge, and $\mathbf{j}$ the applied current density.
Use of this STT expression implies that the conduction
electron spins are always parallel to the magnetic moments
$\mathbf{m}$\cite{iwasaki_universal_2013, zang_dynamics_2011}.
The non-adiabatic terms
can be neglected in this case, since they do not affect the skyrmion dynamics
significantly under small driving forces \cite{litzius_skyrmion_2017}.
The current density $\mathbf{j}$ used in this
work has the form $\mathbf{j}=j\cos\left(2\pi ft\right)\hat{\mathbf{x}}$, where
$j$ is the ac amplitude, $f$ is the ac frequency,
and $t$ is the time.

We fix the following values
in our simulations: $\mu\mathbf{H}=0.6(D^2/J)(-\mathbf{\hat{z}})$,
$\alpha=0.3$, $p=-1.0$, and $f=0.1~$GHz.
The material parameters are $J=1$~meV, $D=0.5J$, and
$K_0=0.1J$.
For each simulation, the system is initialized in one of the
configurations illustrated in Fig.~\ref{fig:1}.
The numerical integration of Eq.~\ref{eq:2} is performed using
a fourth order Runge-Kutta method over 200 ns.

We measure the mean square displacement $R_\alpha$ in each direction, $\alpha=x, y$:
\begin{equation}\label{eq:3}
    R_\alpha(n)=\frac{1}{N_\text{sk}}\sum_i^{N_\text{sk}}\left[\left(\mathbf{r}_i(t_0+nT)-\mathbf{r}_i(t_0)\right)\cdot\hat{\bm{\alpha}}\right]^2
\end{equation}
where $N_\text{sk}$ is the number of skyrmions in the sample, $n$
is the number of ac drive cycles that have elapsed,
$T=1/f=10$~ns is the ac drive period, and $\mathbf{r}_i$ is the position
of skyrmion $i$.
From $R_\alpha$, we compute the effective diffusivity according to
$D_\alpha={R_\alpha}/{2t}$, following the same procedure
used in Refs.~\cite{mangan_reversible_2008, pine_chaos_2005}.

\section{Effective Diffusivity}

\begin{figure}
    \centering
    \includegraphics[width=0.6\columnwidth]{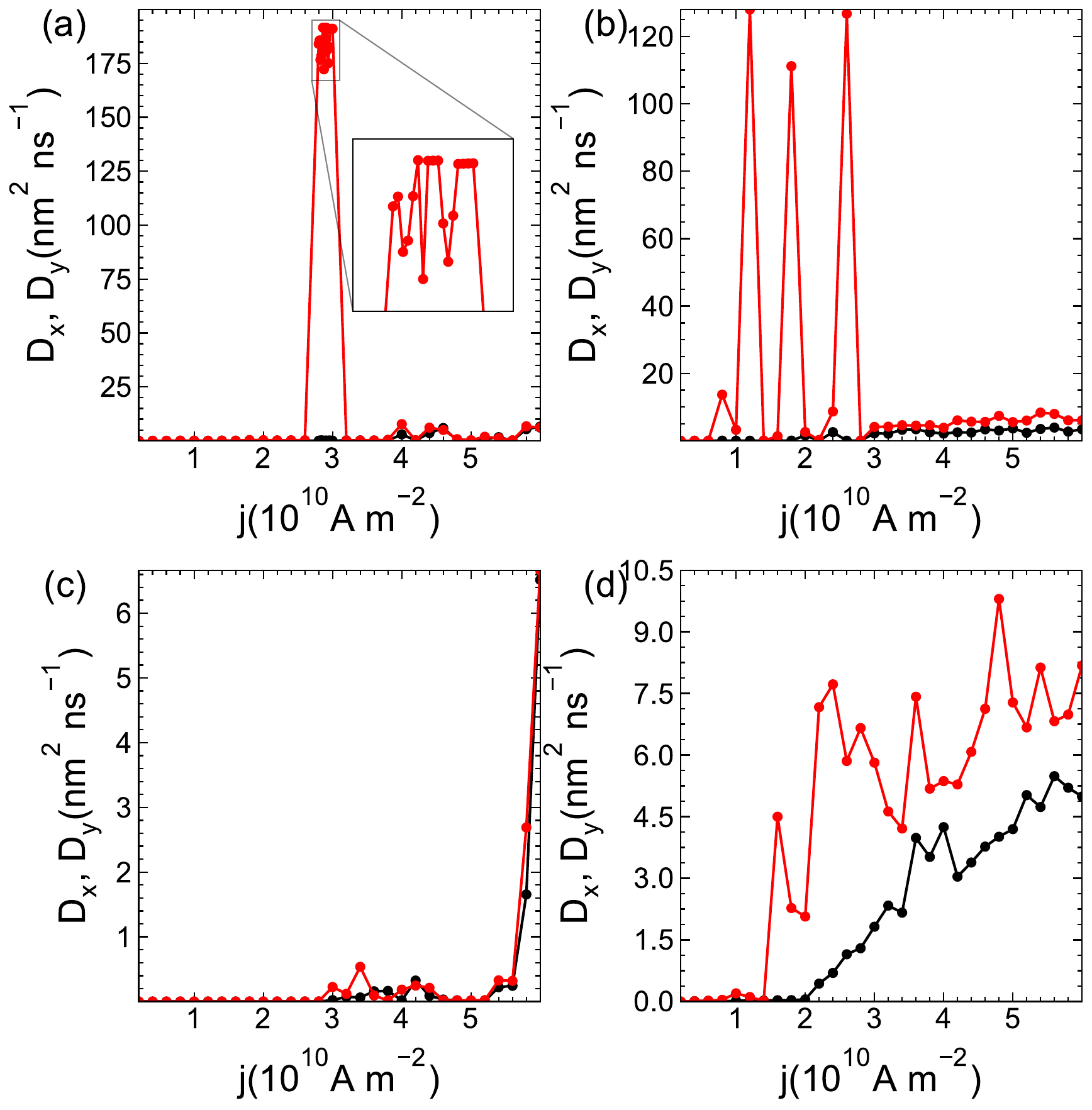}
    \caption{Diffusivities $D_x$ (black) and $D_y$ (red) vs ac
    amplitude $j$ at different fillings.
    (a) $N_{\rm sk}/N_m=2$, the antiferromagnetic dimer arrangement
    from Fig.~\ref{fig:1}(a).
    Inset: a blowup of the region spanning
    $2.7\times10^{10}~\text{A~m}^{-2}\leq j\leq 3.1\times10^{10}~\text{A~m}^{-2}$.
    (b) $N_{\rm sk}/N_m=3$, the alternating column trimer arrangement
    from Fig.~\ref{fig:1}(b).
    (c) $N_{\rm sk}/N_m=3/2$, the bipartite monomer and dimer lattice
    from Fig.~\ref{fig:1}(c).
    (d) $N_{\rm sk}/N_m=5/2$, the bipartite dimer and trimer lattice
    from Fig.~\ref{fig:1}(d).
    Strong
    peaks in $D_y$ appear in panels (a) and (b) for the dimer and trimer
    states, but are absent for both of the bipartite lattices.
    }
    \label{fig:2}
\end{figure}

We first analyze how the effective diffusivity changes
as the filling and ac amplitude $j$ are changed.
As shown in Fig.~\ref{fig:2}(a), for the dimer arrangement
at $N_{\rm sk}/N_m=2$, we obtain
$D_x=0$ and $D_y=0$ for ac amplitudes $j\leq 2.7\times10^{10}~$A~m$^{-2}$.
The absence of diffusion indicates that this is a reversible regime
in which
the skyrmions move back and forth under the ac driving, but return to their
original positions at the end of each driving cycle and experience
no net motion.
An animation of the motion of the skyrmions in this
reversible regime appears in the Supplemental Material
\cite{Suppl}.
Over the interval
$2.8\times10^{10}~\text{A~m}^{-2}\leq j \leq 3.0\times10^{10}~\text{A~m}^{-2}$,
we find that $D_y$ increases abruptly to
$D_y\geq 160$~nm$^2$~ns$^{-1}$ while $D_x$ remains at $D_x=0$,
indicating a net displacement of individual skyrmions only
along the $y$ direction.
When $j$ is increased further,
$D_y$ drops abruptly
back to $D_y=0$.
For $j>3\times10^{10}~$A~m$^{-2}$, $D_x$ and
$D_y$ are very small compared to the peak values
of $D_y$; however,
they are not equal to zero, indicating that the skyrmions are in
an irreversible state with finite net motion
of the skyrmions from cycle to cycle.

We find very different behavior for the $N_{\rm sk}/N_m=3$ trimer
state, as shown in Fig.~\ref{fig:2}(b).
Under very low ac amplitudes, $j\leq 0.8\times10^{10}~$A~m$^{-2}$,
$D_x=D_y=0$. These low ac amplitudes are not
strong enough to dislodge individual skyrmions from the potential
minimum in which they are trapped, producing an oscillation
inside the potential minimum with no net translation.
A small peak in $D_y$ appears for
$j=0.8\times10^{10}~$A~m$^{-2}$, indicating a net motion of the skyrmions
along the $y$ direction.
After this first small peak
at $j=0.8\times10^{10}~$A~m$^{-2}$,
there are three larger peaks in
$D_y$ at $j=1.2\times10^{10}~$A~m$^{-2}$,
$j=1.8\times10^{10}~$A~m$^{-2}$, and $j=2.6\times10^{10}~$A~m$^{-2}$,
with $D_y$ returning to $D_y=0$ between each peak and
$D_x$ remaining at $D_x=0$ throughout the entire current interval.
As was the case for the dimer filling, the peaks in $D_y$ indicate
that there is a net displacement of the skyrmions along
the $y$ direction only.
These peaks have different $D_y$ values,
indicating that different types of motion occur
for each peak. The same is true for the different peak $D_y$ values
in Fig.~\ref{fig:2}(a). 
For higher ac amplitudes of $j\geq 3\times10^{10}~$A~m$^{-2}$,
both $D_x$ and $D_y$ have nonzero values,
indicating a
chaotic irreversible motion of the skyrmions.

For the bipartite monomer and dimer lattice
at $N_{\rm sk}/N_m=3/2$, Fig.~\ref{fig:2}(c) shows that the behavior of
$D_x$ and $D_y$ is much simpler than for the dimer and trimer arrangements.
There are no strong peaks in either $D_x$ or $D_y$,
and over most of the range of ac amplitudes considered here,
$j<3\times10^{10}~$A~m$^{-2}$,
$D_x=0$ and $D_y=0$, indicating that the motion is reversible.
For higher ac amplitude values,
the skyrmions begin
to move in a chaotic irreversible fashion.
The behavior of the bipartite
dimer and trimer lattice at $N_{\rm sk}/N_m=5/2$
in Fig.~\ref{fig:2}(d) is very similar, but has
a reduced ac amplitude interval,
$j\leq 1.4\times10^{10}~$A~m$^{-2}$,
of reversible motion.
The values of $D_x$ and $D_y$ observed in Fig.~\ref{fig:2}(c, d)
in the irreversible state
are similar to the values of $D_x$ and $D_y$ found in
Fig.~\ref{fig:2}(b) for chaotic irreversible motion at
$j\geq 3\times10^{10}~$A~m$^{-2}$.

\section{Dimer Peak Motion}

\begin{figure}
    \centering
    \includegraphics[width=\textwidth]{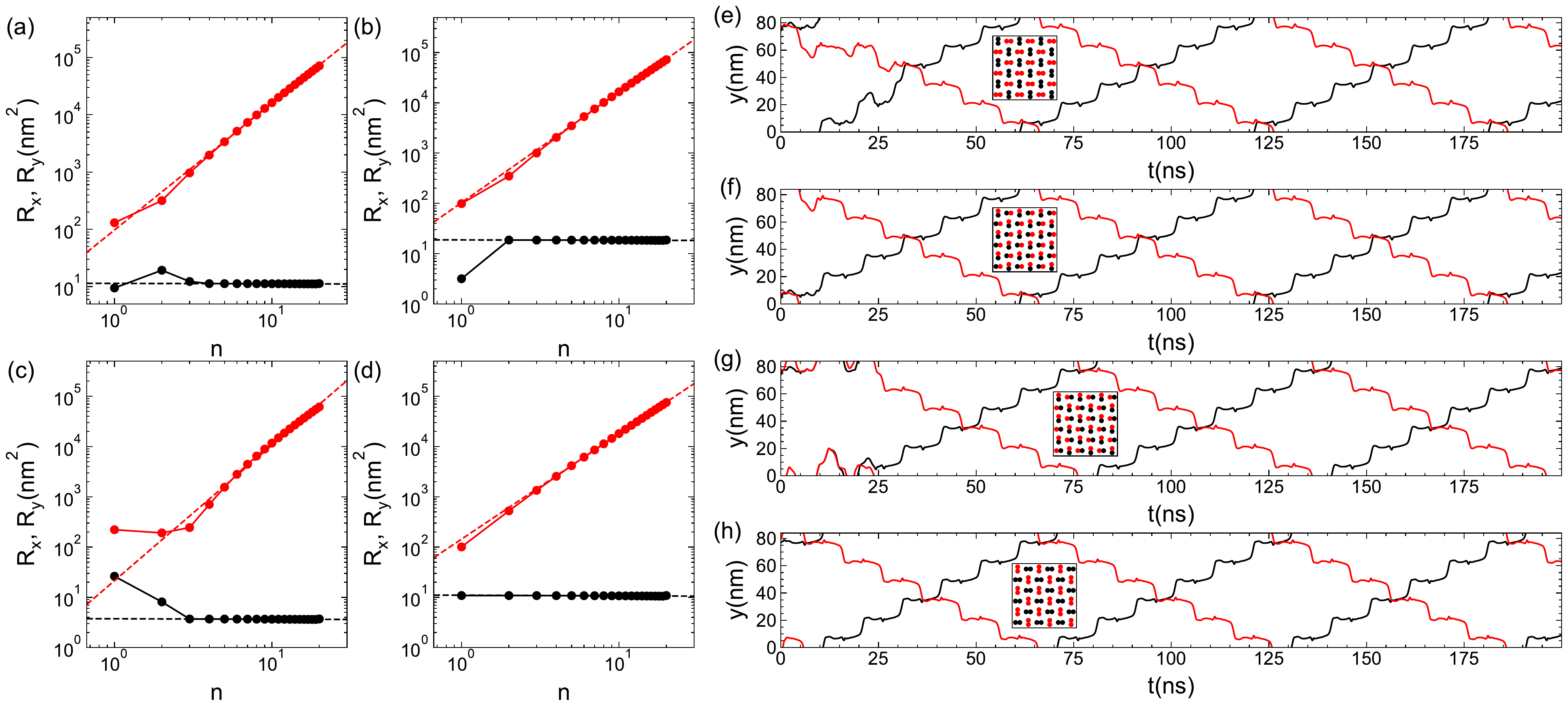}
    \caption{(a-d) $R_x$ (black) and
    $R_y$ (red) vs ac drive cycle number $n$
    for the $N_{\rm sk}/N_m=2$ dimer arrangement from
    Fig.~\ref{fig:1}(a) at ac drive amplitudes of
    (a) $j=2.8\times10^{10}~$A~m$^{-2}$,
    (b) $j=2.85\times10^{10}~$A~m$^{-2}$,
    (c) $j=2.875\times10^{10}~$A~m$^{-2}$, and
    (d) $j=3\times10^{10}~$A~m$^{-2}$.
    Dashed lines are the fitted curves used
    to compute $D_x$ and $D_y$.
    (e-h) The $y$ position of two selected skyrmions versus time
    in samples with
    (e) $j=2.8\times10^{10}~$A~m$^{-2}$,
    (f) $j=2.85\times10^{10}~$A~m$^{-2}$,
    (g) $j=2.875\times10^{10}~$A~m$^{-2}$, and
    (h) $j=3\times10^{10}~$A~m$^{-2}$.
    The curves are colored according to the net motion of each skyrmion:
    $+y$ (black) or $-y$ (red).
    When the skyrmion crosses the periodic boundary, the curve jumps
    from the top to the bottom of the panel or vice versa.
    The insets show the dimer arrangement colored according
    to the net motion of each skyrmion, $+y$ (black) or $-y$ (red).
    In panel (e), all vertically oriented
    dimers have a net $+y$ motion and
    all
    horizontally oriented dimers have a net $-y$ motion.
    In panel (f), the upper skyrmion of each vertically oriented dimer
    moves along $-y$ and the lower skyrmion moves along $+y$, while the
    left skyrmion of each horizontally oriented dimer moves along
    $-y$ and the right skyrmion moves along $+y$.
    In panel (g), the motion is as in panel (f) except that the
    left skyrmion of each horizontally oriented dimer moves along $+y$
    and the right skyrmion moves along $-y$.
    In panel (h) all vertically oriented dimers have a net $-y$ motion
    and all horizontally oriented dimers have a net $+y$ motion.
    An animation showing the skyrmions motion from panels (h) appears
    in the Supplemental Material \cite{Suppl}.
    }
    \label{fig:3}
\end{figure}

We now focus on the dynamical behavior of
the skyrmions at the dimer filling of
$N_{\rm sk}/N_m=2$ in irreversible states with
$D_y\neq 0$.
Figure~\ref{fig:3}(a-d) and Fig.~\ref{fig:3}(e-h)
show the mean square displacement
and individual $y$ trajectories, respectively, for samples at
ac amplitudes of $j=2.8\times10^{10}~$A~m$^{-2}$,
$j=2.85\times10^{10}~$A~m$^{-2}$, $j=2.875\times10^{10}$A~m$^{-2}$, and 
$j=3\times10^{10}~$A~m$^{-2}$, which fall within the region of
large $D_y$ values
in Fig.~\ref{fig:2}(a).
The lowest ac amplitude that produces irreversible behavior
in Fig.~\ref{fig:2}(a)
is $j=2.8\times10^{10}~$A~m$^{-2}$, and we present
$R_x$ and $R_y$ versus ac drive cycle number $n$
for this ac amplitude value
in Fig.~\ref{fig:3}(a). $R_y$ 
increases with $n$ while $R_x$ remains constant.
Both $R_x$ and $R_y$
exhibit a transient behavior for $n<3$, during which the
skyrmions adjust themselves to the presence of the ac driving. This transient
motion is also visible at small times $t<20$ ns
in Fig.~\ref{fig:3}(e), where we plot the time dependence of the $y$
position for two selected skyrmions.
The colors of the curves, black for $+y$ and red for $-y$, indicate the
direction of net motion.
The inset shows
the dimer arrangement with skyrmions colored according to their net
displacement.
After the transient motion period,
all skyrmions from the vertically oriented dimers move along $+y$,
while all skyrmions from the horizontally oriented dimers move
along $-y$.
The $+y$ and $-y$ motions are out of phase by
half of an ac driving period,
since during the $+x$ portion of the ac driving cycle, the
skyrmions from the horizontally oriented dimers act as obstacles
for the skyrmions from the vertically oriented dimers. These roles
reverse during the $-x$ portion of the ac driving cycle.
All of the $+y$ moving skyrmions are in phase with each other,
and all of the $-y$ moving skyrmions are in phase with each other.

For a larger ac drive amplitude of
$j=2.85\times10^{10}~$A~m$^{-2}$,
shown in Fig.~\ref{fig:3}(b, f), the dynamical behavior remains
very similar, but the duration of the transient motion and the
net direction of motion of individual skyrmions changes.
Figure~\ref{fig:3}(b) indicates that
the transient motion appears only for $n<2$, corresponding
to 10 ns as shown in Fig.~\ref{fig:3}(f).
As was the case at the lower value of $j$ above,
half of the skyrmions have a net $+y$ displacement
and the other half have a net $-y$ displacement, but
the spatial distribution of these displacements is different.
For each vertically oriented dimer, the upper skyrmion moves in the $-y$
direction and the lower skyrmion moves in the $+y$ direction. For
each horizontally oriented dimer, the left skyrmion moves in the $+y$
direction and the right skyrmion moves in the $-y$ direction.

Further increasing the ac amplitude to $j=2.875\times10^{10}~$A~m$^{-2}$,
has little impact on the dynamical
behavior, as
shown in Fig.~\ref{fig:3}(c, g).
Now the transient motion extends over the time period $n<3$, as
shown in Fig.~\ref{fig:3}(c), which corresponds to 20 ns in
Fig.~\ref{fig:3}(g).
There is a subtle change in the spatial distribution of the
skyrmion motion
compared to Fig.~\ref{fig:3}(b, f).
For the horizontally oriented dimers, the left skyrmion now moves
in the $-y$ direction instead of the $+y$ direction, and the right
skyrmion moves in the $+y$ direction instead of the $-y$ direction.
This subtle change is responsible for
the increase in the duration of the transient motion.

At an ac amplitude of $j=3\times10^{10}~$A~m$^{-2}$,
illustrated in Fig.~\ref{fig:3}(d, h), the dynamical
behavior changes again. The transient time is almost
zero, meaning that skyrmions start to move with
their steady state net
$y$ displacement after the first ac cycle.
The spatial distribution of skyrmion motion is also different.
All vertically oriented dimers
have a net $-y$ displacement while all horizontally oriented dimers
have a net $+y$ displacement. This is opposite from the motion
found
in Fig.~\ref{fig:3}(e) at $j=2.8\times10^{10}~$A~m$^{-2}$.

For each of the four values of $j$ described above, all of the
skyrmions that are moving along $+y$ travel in phase with each
other, and all of the skyrmions that are moving along $-y$
travel in phase with each other.
Additionally,
the $+y$ and $-y$ motions have the same average velocity
of $\left|\left\langle v_y\right\rangle\right|=1.4$m/s,
which is equal to translation by one substrate
lattice constant per ac cycle, 
$\left|\left\langle v_y\right\rangle\right|=a_0/T$. There is,
however, no net dc motion in the sample, since equal numbers of
skyrmions translate along $+y$ and $-y$.
An animation showing the motion of the skyrmions at $j=3\times10^{10}~$A~m$^{-2}$,
with the same coloring conventions from Fig.~\ref{fig:3}(e-h), appears
in the Supplemental Material \cite{Suppl}.

\section{Trimer Peak Motion}

\begin{figure}
    \centering
    \includegraphics[width=\textwidth]{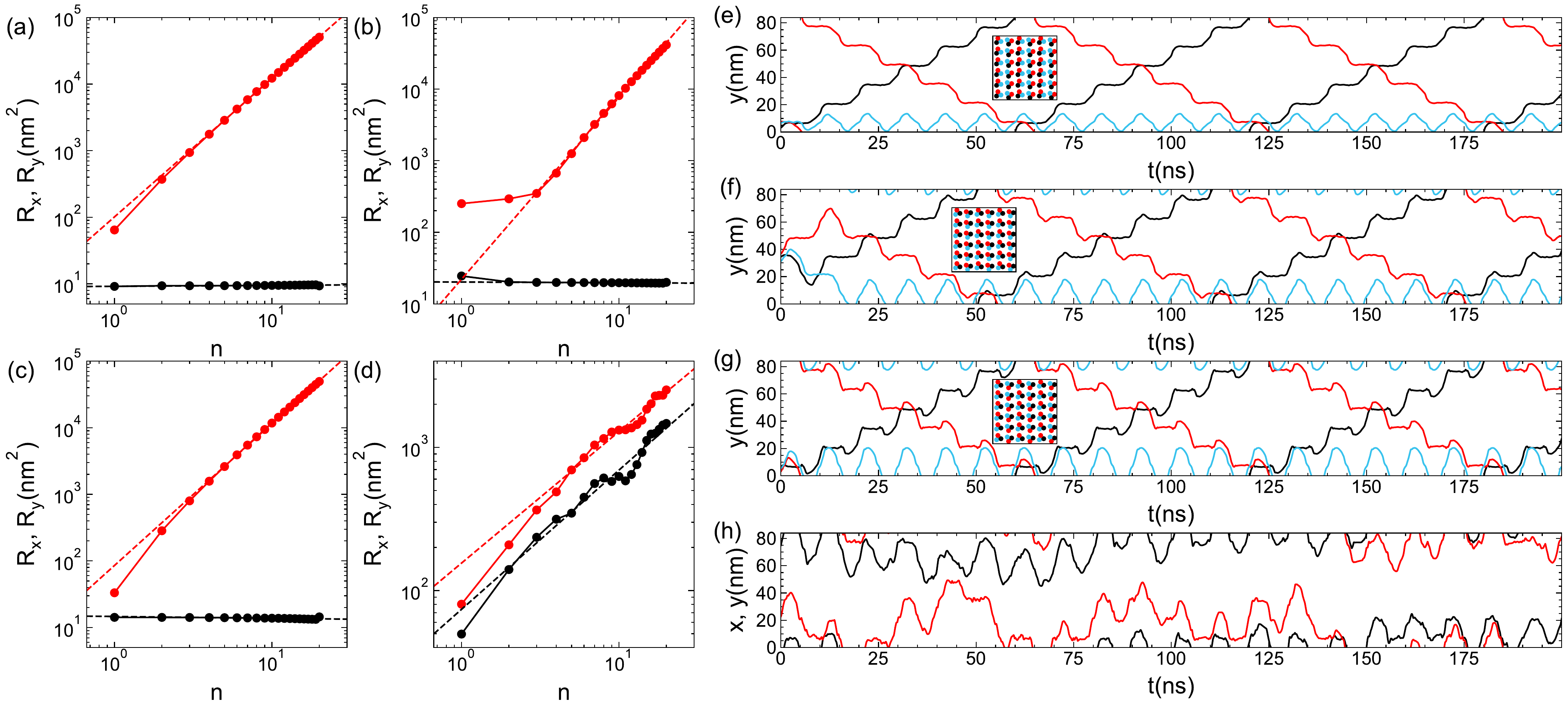}
    \caption{(a-d) $R_x$ (black)
      and $R_y$ (red) vs ac drive cycle number $n$ for the
      $N_{\rm sk}/N_m=3$ trimer
      arrangement from Fig.~\ref{fig:1}(b) at ac
      drive amplitudes of
    (a) $j=1.2\times10^{10}~$A~m$^{-2}$,
    (b) $j=1.8\times10^{10}~$A~m$^{-2}$,
    (c) $j=2.6\times10^{10}~$A~m$^{-2}$, and
    (d) $j=6\times10^{10}~$A~m$^{-2}$. Dashed
      lines are the fitted curves used
      to compute $D_x$ and $D_y$.
      (e-g) The $y$ position of three selected skyrmions versus time in
      samples with 
    (e) $j=1.2\times10^{10}~$A~m$^{-2}$,
    (f) $j=1.8\times10^{10}~$A~m$^{-2}$, and
    (g) $j=2.6\times10^{10}~$A~m$^{-2}$.
      The curves are colored according to the net motion of each skyrmion:
      $+y$ (black), $-y$ (red), and oscillating with no net motion (blue).
      When the skyrmion crosses the periodic
      boundary, the curve jumps from the top to the bottom of the panel or
      vice versa.
      Insets show the spatial distribution of the direction of skyrmion
      motion.
      (h) $x$ (black) and $y$ (red) position of a selected skyrmion versus time
      for a chaotic irreversible system at
    $j=6\times10^{10}~$A~m$^{-2}$.
      An animation of the skyrmions motion from panel(e) appears in the
      Supplemental Material \cite{Suppl}.
      }
    \label{fig:4}
\end{figure}

We now analyze the sharp $D_y$ peaks present
in Fig.~\ref{fig:2}(b) for the
$N_{\rm sk}/N_m=3$
trimer skyrmion
arrangement.
We do not specifically treat the small peak at
    $j=0.8\times10^{10}~$A~m$^{-2}$, since its value
is comparable to the value found for chaotic motion
at larger $j$ and the system exhibits no interesting ordered movement.

The first sharp $D_y$ peak observed in Fig.~\ref{fig:2}(b)
    appears at $j=1.2\times10^{10}~$A~m$^{-2}$, where $D_y\approx130$~nm$^2$~ns$^{-1}$
and $D_x=0$. The corresponding values of $R_x$ and $R_y$ are plotted in
Fig.~\ref{fig:4}(a).
$R_x$ remains constant as $n$ increases,
but $R_y$ increases with increasing $n$, indicating
a net skyrmion displacement along the $y$ direction only.
In Fig.~\ref{fig:4}(e) we plot the $y$ position as a function of time for
three representative skyrmions, one moving along $+y$,
one moving along $-y$, and one oscillating with no net motion.
The oscillating skyrmions
act as an intermediary to
facilitate the $+y$ and $-y$ movement of the other skyrmions.
Skyrmions moving along $+y$ are out of phase by half an
ac drive period from skyrmions moving along $-y$, while
the oscillating skyrmions move in phase with the ac
drive. All skyrmions moving in a particular direction are in phase
with each other, meaning that the $+y$ skyrmions are in phase with
each other, the $-y$ skyrmions are in phase with each other, and the
oscillating skyrmions are in phase with each other.
The transient motion at this
ac amplitude is very brief and lasts only one ac drive cycle, as
seen in Fig.~\ref{fig:4}(a).
The inset of Fig.~\ref{fig:4}(e) indicates the spatial distribution
of the skyrmion motion.
In the odd (first, third, and fifth) columns of trimers, where the trimer tips
are oriented upward, the topmost skyrmion moves along $-y$, the rightmost
skyrmion oscillates, and the leftmost skyrmion moves along $+y$.
In the even (second, fourth, and sixth)
columns of trimers, where the trimer tips
are oriented downward, the bottommost skyrmion moves along $+y$, the
rightmost skyrmion moves along $-y$, and the leftmost skyrmion oscillates.
Viewed clockwise from the $-y$ skyrmion, the order of motion of the
skyrmions in each trimer is $-y$, oscillate, $+y$ for the odd columns and
$-y$, $+y$, oscillate for the even columns.

The second $D_y$ peak occurs at $j=1.8\times10^{10}~$A~m$^{-2}$,
and has a slightly reduced value of $D_y\approx110$~nm$^2$~ns$^{-1}$.
The plots of $R_x$ and $R_y$ for this current, shown
in Fig.~\ref{fig:4}(b), have behavior very similar to that found
for the first peak in $D_y$,
with $R_x$ remaining constant as a function of $n$
and $R_y$ increasing with increasing $n$.
The transient motion lasts for nearly three ac drive cycles,
corresponding to approximately 30 ns in
Fig.~\ref{fig:4}(f).
The dynamical behavior
of the trimers is very similar
what was shown in Fig.~\ref{fig:4}(e),
with each skyrmion exhibiting either $+y$, $-y$, or oscillatory
motion depending on its position inside the trimer. The
spatial distribution of this motion is modified at this
higher value of $j$, as seen by comparing the insets of
Fig.~\ref{fig:4}(e) and Fig.~\ref{fig:4}(f).
In all columns, whether even or odd, the upper leftmost skyrmion
moves along $-y$, the upper rightmost skyrmion moves along
$+y$, and the lower leftmost skyrmion oscillates.
Viewed clockwise from the $-y$ skyrmion, the order of motion of the
skyrmions in all trimers is $-y$, $+y$, oscillate.
The fact that the motion of all trimers synchronizes independent of
whether the trimer is in an odd or even column could be the reason that
the peak value of $D_y$ is somewhat diminished and the transient time
is extended.

The last $D_y$ peak for the trimer arrangement falls at
$j=2.6\times10^{10}~$A~m$^{-2}$, where we find
$D_y\approx130$~nm$^2$~ns$^{-1}$, a higher value than what appears
for $j=1.8\times10^{10}~$A~m$^{-2}$.
The corresponding $R_x$ and
$R_y$ versus $n$ curves plotted
in Fig.~\ref{fig:4}(c) are very similar to
those found for the other $D_y$ peaks,
with $R_x$ remaining constant
and $R_y$ increasing with the
number of ac drive cycles.
The transient time at this
ac amplitude is reduced compared to the transient
time observed in Fig.~\ref{fig:4}(b), and lasts for only
one ac drive cycle, corresponding to $t<10$ ns in
Fig.~\ref{fig:4}(g).
As with the other peaks in $D_y$, for the dynamical behavior
of the skyrmions we find that each skyrmion either moves in the $+y$ or
$-y$ direction or oscillates about a fixed point.
The spatial distribution of the skyrmion motion,
in the inset of Fig.~\ref{fig:4}(g), is different
from that at the other peaks in $D_y$.
Trimers in odd columns exhibit the same ordering found in Fig.~\ref{fig:4}(f),
with the topmost skyrmion moving along $-y$, the lower rightmost skyrmion
moving along $+y$, and the lower leftmost skyrmion oscillating.
Trimers in even columns have an ordering different from that found for
other values of $j$, with the bottommost skyrmion moving along $-y$,
the upper leftmost skyrmion oscillating, and the upper rightmost skyrmion
moving along $+y$.
Viewed clockwise from the $-y$ skyrmion, the order of motion of the
skyrmions is $-y$, $+y$, oscillate for the odd columns and
$-y$, oscillate, $+y$ for the even columns.
In both Fig.~\ref{fig:4}(e) and Fig.~\ref{fig:4}(g),
where the order of trimer motion reverses from one column to the
next,
the $D_y$ peak values are higher compared to the case in Fig.~\ref{fig:4}(f)
where all columns have the same order of trimer motion.
The average skyrmion displacement velocities are, however,
the same in all cases, $\left|\left\langle v_y\right\rangle\right|=a_0/T$,
meaning that all skyrmions
that undergo a net displacement translate by one substrate lattice
constant per ac drive period.
This is the same velocity observed for the dimer system.

Figure~\ref{fig:4}(d) shows the values of $R_x$ and $R_y$ versus $n$ for
the trimer system at $j=6\times10^{10}~$A~m$^{-2}$.
At this ac amplitude the skyrmions do not exhibit an ordered motion,
but instead undergo an irreversible chaotic motion.
Both $R_x$ and $R_y$ increase
with increasing $n$.
This is very different from the behavior at peak values of $D_y$,
where $R_x$ remained constant while $R_y$ increased
with increasing $n$.
The values of $R_x$ and $R_y$ for the disordered or chaotic irreversible
motion
are two orders of magnitude smaller than the value of $R_y$ that appears for
ordered irreversible motion.
This reduction can be seen
in Fig.~\ref{fig:2}(b),
where for ac amplitudes $j\geq 3\times10^{10}~$A~m$^{-2}$,
the skyrmions exhibit an irreversible chaotic motion with finite $D_x$ and
$D_y$ that have values much smaller than the peak values of $D_y$ that
appear for the ordered irreversible motion.
The $x$ and $y$ positions of a representative skyrmion are
plotted as a function of time at $j=6\times10^{10}~$A~m$^{-2}$
in Fig.~\ref{fig:4}(h). The skyrmion motion is chaotic, and the
displacement velocity is greatly reduced compared to the velocity that
appears for the irreversible ordered motion.
An animation of the motion of the trimers at $j=3\times10^{10}~$A~m$^{-2}$ appears
in the Supplemental Material \cite{Suppl}.

\begin{figure}
    \centering
    \includegraphics[width=0.6\columnwidth]{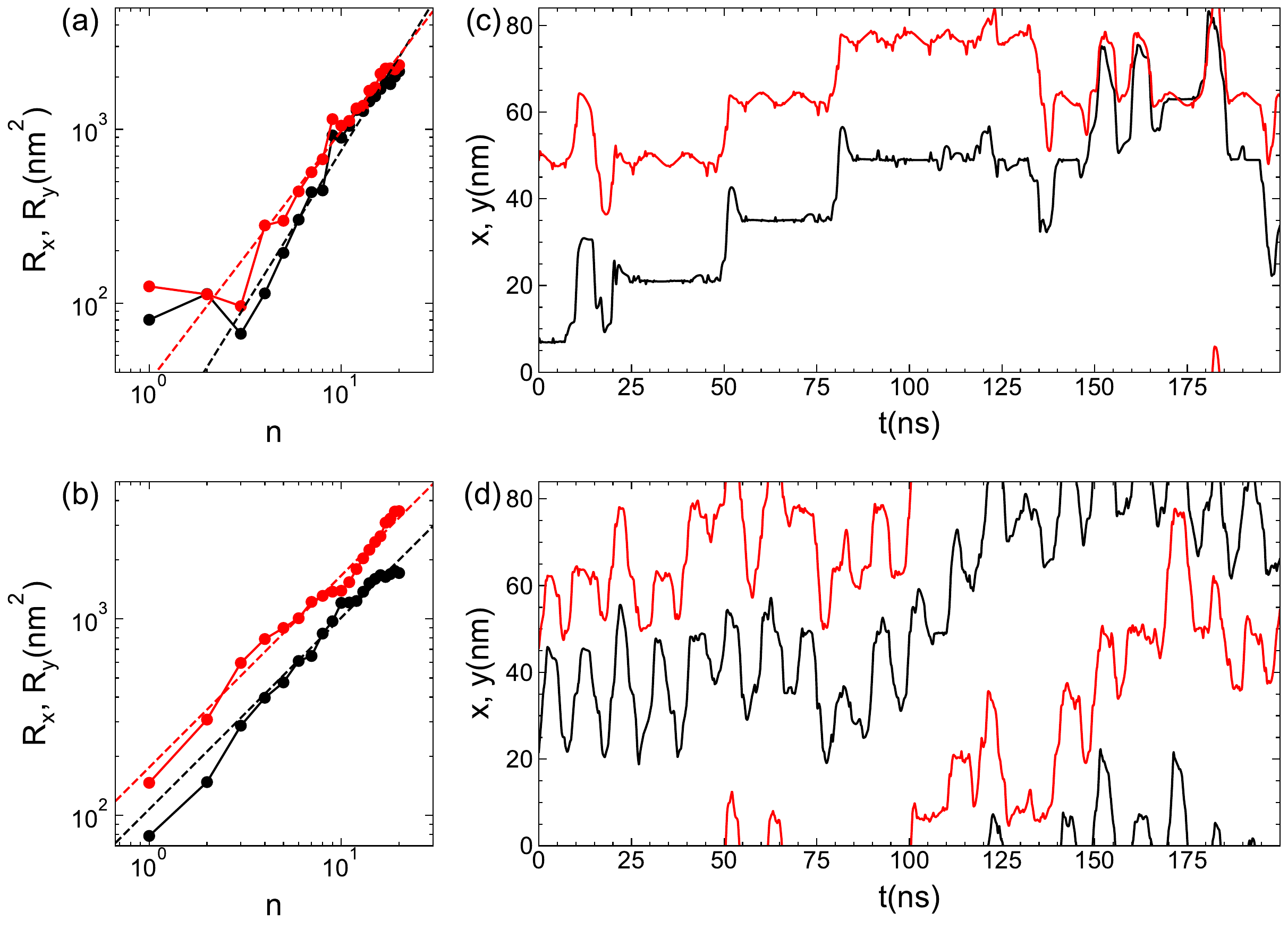}
    \caption{(a,b) $R_x$ (black) and
      $R_y$ (red) vs ac drive cycle number $n$
    at $j=6\times10^{10}~$A~m$^{-2}$
      for 
    (a) the $N_{\rm sk}/N_m=3/2$ bipartite monomer and dimer lattice and
    (b) the $N_{\rm sk}/N_m=5/2$ bipartite dimer and trimer lattice.
      Dashed lines are the fitted curves used
      to compute $D_x$ and $D_y$.
      (c,d) $x$ (black) and $y$ (red) position of a selected skyrmion
      versus time at
    $j=6\times10^{10}~$A~m$^{-2}$ in
    (c) the $N_{\rm sk}/N_m=3/2$ bipartite monomer and dimer lattice and
    (d) the $N_{\rm sk}/N_m=5/2$ bipartite dimer and trimer lattice.
    Animations showing the skyrmion motion are
    available in the Supplemental Material \cite{Suppl}.
    }
    \label{fig:5}
\end{figure}

\section{Bipartite Lattices}

Finally, we analyze the
$N_{\rm sk}/N_m=3/2$
bipartite monomer and dimer lattice and the
$N_{\rm sk}/N_m=5/2$
bipartite dimer and trimer lattice
in Fig.~\ref{fig:5}.
The low values of $D_x$ and $D_y$ in
Fig.~\ref{fig:2}(c,d) for both bipartite lattices indicates that
neither undergoes
any type of ordered motion, but both show only
chaotic irreversible motion.
The $N_{\rm sk}/N_m=3/2$
bipartite lattice first enters the chaotic motion regime
at large ac driving amplitudes,
while
the $N_{\rm sk}/N_m=5/2$ bipartite lattice
already exhibits chaotic
motion under low ac drive amplitudes.

The $R_x$ and $R_y$ versus $n$ curves for the
$N_{\rm sk}/N_m=3/2$ bipartite lattice
in Fig.~\ref{fig:5}(a) both increase with increasing $n$.
This is the expected behavior for chaotic motion, and matches what
was observed in the trimer system
under large ac drive amplitudes
in Fig.~\ref{fig:4}(d).
The values of $R_x$ and $R_y$ are
two to three orders of magnitude smaller than
the $R_y$ values that appear
during ordered motion, such as in Fig.~\ref{fig:4}(a-c).
We illustrate the chaotic motion in
Fig.~\ref{fig:5}(c) by plotting the $x$ and $y$ positions of a selected
skyrmion as a function of time.
The skyrmion motion has 
no well defined behavior, and the skyrmion diffuses through
the sample over time.
The same dynamical behavior is observed for the
$N_{\rm sk}/N_m=5/2$ bipartite lattice, as shown
in Fig.~\ref{fig:5}(b), where
both $R_x$ and $R_y$ increase with increasing $n$, but
maintain values that are
two to three orders of magnitude smaller than the
values of $R_y$ found for ordered motion.
The $x$ and $y$ positions of a selected skyrmion versus time,
plotted in Fig.~\ref{fig:5}(d), indicate the
presence of chaotic motion. There is greater variability in the
motion as a function of time compared to
Fig.~\ref{fig:5}(c) due to the higher total filling fraction
in the 
$N_{\rm sk}/N_m=5/2$ bipartite lattice,
which lowers the
effective depinning threshold and enables the skyrmions
to diffuse around the sample with higher mobility.
Animations showing the chaotic skyrmion motion for
the bipartite
lattices at $j=6\times10^{10}~$A~m$^{-2}$ are available in the
Supplemental Material \cite{Suppl}.

\section{Summary}

Using atomistic simulations,
we investigated the reversible and irreversible behavior
of ac driven skyrmions interacting with a periodic substrate for
dimer, trimer, bipartite monomer and dimer, and bipartite dimer and trimer
fillings.
We find that under low ac amplitudes, all of the systems exhibit
reversible dynamics in which all skyrmions in the sample return to their
starting positions and the long time diffusion both parallel and
perpendicular to the applied ac drive is zero.
At larger ac amplitudes for the dimer and trimer
systems, we observe pronounced peaks in the perpendicular
diffusion $D_y$ while the parallel diffusion $D_x$ remains zero,
indicating that individual skyrmions are undergoing
one-dimensional directed motion along the $y$ direction.
In this ordered irreversible state,
the skyrmions
move one lattice constant per ac drive cycle in the translating regimes.
For the dimers, half of the skyrmions move along $+y$ and the other half
move along $-y$,
while for the trimers, a third of the skyrmions move along $+y$, a third
move along $-y$, and the remaining third undergo confined oscillatory
motion, so that there is no net skyrmion motion for either the dimer or
trimer system.
Under large ac drive amplitudes, we find
irreversible chaotic motion of the dimers, trimers, and
bipartite lattices.
During the chaotic motion,
the skyrmions diffuse throughout the sample in both the $x$ and
$y$ directions.
In the chaotic phases, the displacements grow diffusively,
while in the translating
phase, the displacements grow ballistically.
Our results suggest a very precise way of controlling the skyrmion
motion under ac driving for different skyrmion arrangements. The
constant ballistic velocity of one substrate lattice
constant per ac period can
be exploited to construct spintronic devices in which the skyrmion
motion and velocity must be carefully regulated.

\section*{Acknowledgments}
We gratefully acknowledge the support of the U.S. Department of
Energy through the LANL/LDRD program for this work.
This work was supported by the US Department of Energy through the Los Alamos National Laboratory. Los
Alamos National Laboratory is operated by Triad National Security, LLC, for the National Nuclear Security
Administration of the U. S. Department of Energy (Contract No. 892333218NCA000001). 
J.C.B.S acknowledges funding from Fundação de Amparo à Pesquisa do Estado de São Paulo - FAPESP (Grant 2023/17545-1).
We would like to thank Dr. Felipe F. Fanchini for providing the computational resources used in this work. 
These resources were funded by the Fundação de Amparo à Pesquisa do Estado de São Paulo - FAPESP (Grant: 2021/04655-8).

\section*{References}
\bibliographystyle{unsrt}
\bibliography{mybib}

\begin{thebibliography}{10}

\bibitem{reichhardt_depinning_2017}
C.~Reichhardt and C.~J.~Olson Reichhardt.
\newblock Depinning and nonequilibrium dynamic phases of particle assemblies
  driven over random and ordered substrates: a review.
\newblock {\em Rep. Prog. Phys.}, 80(2):26501, 2017.

\bibitem{pertsinidis_statics_2008}
A.~Pertsinidis and X.~S. Ling.
\newblock Statics and dynamics of {2D} colloidal crystals in a random pinning
  potential.
\newblock {\em Phys. Rev. Lett.}, 100(2):028303, 2008.

\bibitem{tierno_depinning_2012}
P.~Tierno.
\newblock Depinning and collective dynamics of magnetically driven colloidal
  monolayers.
\newblock {\em Phys. Rev. Lett.}, 109(19):198304, 2012.

\bibitem{bhattacharya_dynamics_1993}
S.~Bhattacharya and M.~J. Higgins.
\newblock Dynamics of a disordered flux line lattice.
\newblock {\em Phys. Rev. Lett.}, 70(17):2617--2620, 1993.

\bibitem{shaw_critical_2012}
G.~Shaw, P.~Mandal, S.~S. Banerjee, A.~Niazi, A.~K. Rastogi, A.~K. Sood,
  S.~Ramakrishnan, and A.~K. Grover.
\newblock Critical behavior at depinning of driven disordered vortex matter in
  2{H-NbS}$_2$.
\newblock {\em Phys. Rev. B}, 85:174517, 2012.

\bibitem{bechinger_active_2016}
C.~Bechinger, R.~Di~Leonardo, H.~L\"owen, C.~Reichhardt, G.~Volpe, and
  G.~Volpe.
\newblock Active particles in complex and crowded environments.
\newblock {\em Rev. Mod. Phys.}, 88:045006, 2016.

\bibitem{vanossi_colloquium_2013}
A.~Vanossi, N.~Manini, M.~Urbakh, S.~Zapperi, and E.~Tosatti.
\newblock Colloquium: {M}odeling friction: From nanoscale to mesoscale.
\newblock {\em Rev. Mod. Phys.}, 85:529--552, 2013.

\bibitem{carlson_properties_1989}
J.~M. Carlson and J.~S. Langer.
\newblock Properties of earthquakes generated by fault dynamics.
\newblock {\em Phys. Rev. Lett.}, 62(22):2632--2635, 1989.

\bibitem{pine_chaos_2005}
D.~J. Pine, J.~P. Gollub, J.~F. Brady, and A.~M. Leshansky.
\newblock Chaos and threshold for irreversibility in sheared suspensions.
\newblock {\em Nature (London)}, 438(7070):997--1000, 2005.

\bibitem{corte_random_2008}
L.~Cort{\' e}, P.~M. Chaikin, J.~P. Gollub, and D.~J. Pine.
\newblock Random organization in periodically driven systems.
\newblock {\em Nature Phys.}, 4(5):420--424, 2008.

\bibitem{schreck_particle-scale_2013}
C.~F. Schreck, R.~S. Hoy, M.~D. Shattuck, and C.~S. O’Hern.
\newblock Particle-scale reversibility in athermal particulate media below
  jamming.
\newblock {\em Phys. Rev. E}, 88(5):052205, 2013.

\bibitem{milz_connecting_2013}
L.~Milz and M.~Schmiedeberg.
\newblock Connecting the random organization transition and jamming within a
  unifying model system.
\newblock {\em Phys. Rev. E}, 88(6):062308, 2013.

\bibitem{zhou_random_2014}
C.~Zhou, C.~J. Olson~Reichhardt, C.~Reichhardt, and I.~Beyerlein.
\newblock Random organization in periodically driven gliding dislocations.
\newblock {\em Phys. Lett. A}, 378(22):1675--1678, 2014.

\bibitem{ni_yield_2019}
X.~Ni, H.~Zhang, D.~B. Liarte, L.~W. McFaul, K.~A. Dahmen, J.~P. Sethna, and
  J.~R. Greer.
\newblock Yield precursor dislocation avalanches in small crystals: The
  irreversibility transition.
\newblock {\em Phys. Rev. Lett.}, 123(3):035501, 2019.

\bibitem{regev_onset_2013}
I.~Regev, T.~Lookman, and C.~Reichhardt.
\newblock Onset of irreversibility and chaos in amorphous solids under periodic
  shear.
\newblock {\em Phys. Rev. E}, 88(6):062401, 2013.

\bibitem{regev_reversibility_2015}
I.~Regev, J.~Weber, C.~Reichhardt, K.~A. Dahmen, and T.~Lookman.
\newblock Reversibility and criticality in amorphous solids.
\newblock {\em Nature Commun.}, 6(1):8805, 2015.

\bibitem{priezjev_reversible_2016}
N.~V. Priezjev.
\newblock Reversible plastic events during oscillatory deformation of amorphous
  solids.
\newblock {\em Phys. Rev. E}, 93(1):013001, 2016.

\bibitem{leishangthem_yielding_2017}
P.~Leishangthem, A.~D.~S. Parmar, and S.~Sastry.
\newblock The yielding transition in amorphous solids under oscillatory shear
  deformation.
\newblock {\em Nature Commun.}, 8(1):14653, 2017.

\bibitem{jana_irreversibility_2017}
P.~K. Jana, M.~J. Alava, and S.~Zapperi.
\newblock Irreversibility transition of colloidal polycrystals under cyclic
  deformation.
\newblock {\em Sci. Rep.}, 7(1):45550, 2017.

\bibitem{mangan_reversible_2008}
N.~Mangan, C.~Reichhardt, and C.~J.~Olson Reichhardt.
\newblock Reversible to irreversible flow transition in periodically driven
  vortices.
\newblock {\em Phys. Rev. Lett.}, 100(18):187002, 2008.

\bibitem{okuma_transition_2011}
S.~Okuma, Y.~Tsugawa, and A.~Motohashi.
\newblock Transition from reversible to irreversible flow: Absorbing and
  depinning transitions in a sheared-vortex system.
\newblock {\em Phys. Rev. B}, 83(1):012503, 2011.

\bibitem{dobroka_memory_2017}
M.~Dobroka, Y.~Kawamura, K.~Ienaga, S.~Kaneko, and S.~Okuma.
\newblock Memory formation and evolution of the vortex configuration associated
  with random organization.
\newblock {\em New J. Phys.}, 19(5):053023, 2017.

\bibitem{brown_reversible_2019}
B.~L. Brown, C.~Reichhardt, and C.~J.~O. Reichhardt.
\newblock Reversible to irreversible transitions in periodically driven
  skyrmion systems.
\newblock {\em New J. Phys.}, 21(1):013001, 2019.

\bibitem{Reichhardt23}
C.~Reichhardt, Ido Regev, K.~Dahmen, S.~Okuma, and C.~J.~O. Reichhardt.
\newblock Reversible to irreversible transitions in periodic driven many-body
  systems and future directions for classical and quantum systems.
\newblock {\em Phys. Rev. Res.}, 5:021001, 2023.

\bibitem{Harada96}
K.~Harada, O.~Kamimura, H.~Kasai, T.~Matsuda, A.~Tonomura, and V.~V.
  Moshchalkov.
\newblock Direct observation of vortex dynamics in superconducting films with
  regular arrays of defects.
\newblock {\em Science}, 274(5290):1167--1170, 1996.

\bibitem{reichhardt_commensurate_1998}
C.~Reichhardt, C.~J. Olson, and F.~Nori.
\newblock Commensurate and incommensurate vortex states in superconductors with
  periodic pinning arrays.
\newblock {\em Phys. Rev. B}, 57:7937--7943, 1998.

\bibitem{Bohlein12}
T.~Bohlein, J.~Mikhael, and C.~Bechinger.
\newblock Observation of kinks and antikinks in colloidal monolayers driven
  across ordered surfaces.
\newblock {\em Nature Mater.}, 11(2):126--130, 2012.

\bibitem{nagaosa_topological_2013}
N.~Nagaosa and Y.~Tokura.
\newblock Topological properties and dynamics of magnetic skyrmions.
\newblock {\em Nature Nanotechnol.}, 8(12):899--911, December 2013.

\bibitem{je_direct_2020}
S.-G. Je, H.-S. Han, S.~K. Kim, S.~A. Montoya, W.~Chao, I.-S. Hong, E.~E.
  Fullerton, K.-S. Lee, K.-J. Lee, M.-Y. Im, and J.-I. Hong.
\newblock Direct demonstration of topological stability of magnetic skyrmions
  \textit{via} topology manipulation.
\newblock {\em ACS Nano}, 14(3):3251--3258, March 2020.

\bibitem{olson_reichhardt_comparing_2014}
C.~J. Olson~Reichhardt, S.~Z. Lin, D.~Ray, and C.~Reichhardt.
\newblock Comparing the dynamics of skyrmions and superconducting vortices.
\newblock {\em Physica C}, 503:52--57, 2014.

\bibitem{reichhardt_statics_2022}
C.~Reichhardt, C.~J.~O. Reichhardt, and M.~Milo{\v s}evi{\' c}.
\newblock Statics and dynamics of skyrmions interacting with disorder and
  nanostructures.
\newblock {\em Rev. Mod. Phys.}, 94:035005, 2022.

\bibitem{litzius_skyrmion_2017}
K.~Litzius, I.~Lemesh, B.~Kr{\" u}ger, P.~Bassirian, L.~Caretta, K.~Richter,
  F.~B{\" u}ttner, K.~Sato, O.~A. Tretiakov, J.~F{\" o}rster, R.~M. Reeve,
  M.~Weigand, I.~Bykova, H.~Stoll, G.~Sch{\" u}tz, G.~S.~D. Beach, and M.~Kl{\"
  a}ui.
\newblock Skyrmion {Hall} effect revealed by direct time-resolved {X}-ray
  microscopy.
\newblock {\em Nature Phys.}, 13(2):170--175, 2017.

\bibitem{iwasaki_universal_2013}
J.~Iwasaki, M.~Mochizuki, and N.~Nagaosa.
\newblock Universal current-velocity relation of skyrmion motion in chiral
  magnets.
\newblock {\em Nature Commun.}, 4(1):1463, February 2013.

\bibitem{jiang_direct_2017}
W.~Jiang, X.~Zhang, G.~Yu, W.~Zhang, X.~Wang, M.~B. Jungfleisch, J.~E. Pearson,
  X.~Cheng, O.~Heinonen, K.~L. Wang, Y.~Zhou, A.~Hoffmann, and S.~G.~E.
  te~Velthuis.
\newblock Direct observation of the skyrmion {Hall} effect.
\newblock {\em Nature Phys.}, 13(2):162--169, 2017.

\bibitem{lin_driven_2013}
S.-Z. Lin, C.~Reichhardt, C.~D. Batista, and A.~Saxena.
\newblock Driven skyrmions and dynamical transitions in chiral magnets.
\newblock {\em Phys. Rev. Lett.}, 110(20):207202, May 2013.

\bibitem{lin_particle_2013}
S.-Z. Lin, C.~Reichhardt, C.~D. Batista, and A.~Saxena.
\newblock Particle model for skyrmions in metallic chiral magnets: Dynamics,
  pinning, and creep.
\newblock {\em Phys. Rev. B}, 87(21):214419, June 2013.

\bibitem{feilhauer_controlled_2020}
J.~Feilhauer, S.~Saha, J.~Tobik, M.~Zelent, L.~J. Heyderman, and
  M.~Mruczkiewicz.
\newblock Controlled motion of skyrmions in a magnetic antidot lattice.
\newblock {\em Phys. Rev. B}, 102(18):184425, November 2020.

\bibitem{reichhardt_quantized_2015}
C.~Reichhardt, D.~Ray, and C.~J.~Olson Reichhardt.
\newblock Quantized transport for a skyrmion moving on a two-dimensional
  periodic substrate.
\newblock {\em Phys. Rev. B}, 91(10):104426, March 2015.

\bibitem{reichhardt_nonequilibrium_2018}
C.~Reichhardt, D.~Ray, and C.~J.~O. Reichhardt.
\newblock Nonequilibrium phases and segregation for skyrmions on periodic
  pinning arrays.
\newblock {\em Phys. Rev. B}, 98(13):134418, October 2018.

\bibitem{vizarim_directional_2021}
N.~P. Vizarim, J.~C.~Bellizotti Souza, C.~Reichhardt, C.~J.~O. Reichhardt, and
  P.~A. Venegas.
\newblock Directional locking and the influence of obstacle density on skyrmion
  dynamics in triangular and honeycomb arrays.
\newblock {\em J. Phys: Condens. Matter}, 33(30):305801, June 2021.

\bibitem{vizarim_skyrmion_2020}
N.~P. Vizarim, C.~Reichhardt, C.~J.~O. Reichhardt, and P.~A. Venegas.
\newblock Skyrmion dynamics and topological sorting on periodic obstacle
  arrays.
\newblock {\em New J. Phys.}, 22(5):53025, May 2020.

\bibitem{reichhardt_commensuration_2022}
C.~Reichhardt and C.~J.~O. Reichhardt.
\newblock Commensuration effects on skyrmion {Hall} angle and drag for
  manipulation of skyrmions on two-dimensional periodic substrates.
\newblock {\em Phys. Rev. B}, 105(21):214437, 2022.

\bibitem{vizarim_guided_2021}
N.~P. Vizarim, C.~Reichhardt, P.~A. Venegas, and C.~J.~O. Reichhardt.
\newblock Guided skyrmion motion along pinning array interfaces.
\newblock {\em J. Mag. Mag. Mater.}, 528:167710, June 2021.

\bibitem{zhang_edge-guided_2022}
C.-L. Zhang, J.-N. Wang, C.-K. Song, N.~Mehmood, Z.-Z. Zeng, Y.-X. Ma, J.-B.
  Wang, and Q.-F. Liu.
\newblock Edge-guided heart-shaped skyrmion.
\newblock {\em Rare Metals}, 41(3):865--870, March 2022.

\bibitem{reichhardt_magnus-induced_2015}
C.~Reichhardt, D.~Ray, and C.~J.~Olson Reichhardt.
\newblock Magnus-induced ratchet effects for skyrmions interacting with
  asymmetric substrates.
\newblock {\em New J. Phys.}, 17(7):73034, July 2015.

\bibitem{souza_skyrmion_2021}
J.~C.~Bellizotti Souza, N.~P. Vizarim, C.~J.~O. Reichhardt, C.~Reichhardt, and
  P.~A. Venegas.
\newblock Skyrmion ratchet in funnel geometries.
\newblock {\em Phys. Rev. B}, 104(5):54434, August 2021.

\bibitem{chen_skyrmion_2019}
W.~Chen, L.~Liu, Y.~Ji, and Y.~Zheng.
\newblock Skyrmion ratchet effect driven by a biharmonic force.
\newblock {\em Phys. Rev. B}, 99(6):64431, February 2019.

\bibitem{gobel_skyrmion_2021}
B.~G{\" o}bel and I.~Mertig.
\newblock Skyrmion ratchet propagation: utilizing the skyrmion {Hall} effect in
  {AC} racetrack storage devices.
\newblock {\em Sci. Rep.}, 11(1):3020, February 2021.

\bibitem{souza_controlled_2024}
J.~C.~Bellizotti Souza, N.~P. Vizarim, C.~J.~O. Reichhardt, C.~Reichhardt, and
  P.~A. Venegas.
\newblock Controlled skyrmion ratchet in linear protrusion defects.
\newblock {\em Phys. Rev. B}, 109(5):054407, 2024.

\bibitem{yanes_skyrmion_2019}
R.~Yanes, F.~Garcia-Sanchez, R.~F. Luis, E.~Martinez, V.~Raposo, L.~Torres, and
  L.~Lopez-Diaz.
\newblock Skyrmion motion induced by voltage-controlled in-plane strain
  gradients.
\newblock {\em Appl. Phys. Lett.}, 115(13):132401, September 2019.

\bibitem{zhang_manipulation_2018}
S.~L. Zhang, W.~W. Wang, D.~M. Burn, H.~Peng, H.~Berger, A.~Bauer,
  C.~Pfleiderer, G.~van~der Laan, and T.~Hesjedal.
\newblock Manipulation of skyrmion motion by magnetic field gradients.
\newblock {\em Nature Commun.}, 9(1):2115, May 2018.

\bibitem{everschor_rotating_2012}
K.~Everschor, M.~Garst, B.~Binz, F.~Jonietz, S.~M{\" u}hlbauer, C.~Pfleiderer,
  and A.~Rosch.
\newblock Rotating skyrmion lattices by spin torques and field or temperature
  gradients.
\newblock {\em Phys. Rev. B}, 86(5):54432, August 2012.

\bibitem{kong_dynamics_2013}
L.~Kong and J.~Zang.
\newblock Dynamics of an insulating skyrmion under a temperature gradient.
\newblock {\em Phys. Rev. Lett.}, 111(6):67203, August 2013.

\bibitem{gong_current-driven_2020}
X.~Gong, H.~Y. Yuan, and X.~R. Wang.
\newblock Current-driven skyrmion motion in granular films.
\newblock {\em Phys. Rev. B}, 101(6):064421, February 2020.

\bibitem{del-valle_defect_2022}
N.~Del-Valle, J.~Castell-Queralt, L.~Gonz{\' a}lez-G{\' o}mez, and C.~Navau.
\newblock Defect modeling in skyrmionic ferromagnetic systems.
\newblock {\em APL Mater.}, 10:10702, 2022.

\bibitem{yuan_wiggling_2019}
H.~Y. Yuan, X.~S. Wang, Man-Hong Yung, and X.~R. Wang.
\newblock Wiggling skyrmion propagation under parametric pumping.
\newblock {\em Phys. Rev. B}, 99(1):014428, January 2019.

\bibitem{zhang_magnetic_2015}
X.~Zhang, Y.~Zhou, M.~Ezawa, G.~P. Zhao, and W.~Zhao.
\newblock Magnetic skyrmion transistor: skyrmion motion in a voltage-gated
  nanotrack.
\newblock {\em Sci. Rep.}, 5(1):11369, June 2015.

\bibitem{zhao_ferromagnetic_2020}
L.~Zhao, X.~Liang, J.~Xia, G.~Zhao, and Y.~Zhou.
\newblock A ferromagnetic skyrmion-based diode with a voltage-controlled
  potential barrier.
\newblock {\em Nanoscale}, 17:9507, 2020.

\bibitem{zhang_structural_2022}
X.~Zhang, J.~Xia, and X.~Liu.
\newblock Structural transition of skyrmion quasiparticles under compression.
\newblock {\em Phys. Rev. B}, 105(18):184402, 2022.

\bibitem{bellizotti_souza_spontaneous_2023}
J.~C. Bellizotti~Souza, N.~P. Vizarim, C.~J.~O. Reichhardt, C.~Reichhardt, and
  P.~A. Venegas.
\newblock Spontaneous skyrmion conformal lattice and transverse motion during
  dc and ac compression.
\newblock {\em New J. Phys.}, 25(5):053020, May 2023.

\bibitem{vizarim_soliton_2022}
N.~P. Vizarim, J.~C.~Bellizotti Souza, C.~J.~O. Reichhardt, C.~Reichhardt,
  M.~V. Milo{\v s}evi{\' c}, and P.~A. Venegas.
\newblock Soliton motion in skyrmion chains: Stabilization and guidance by
  nanoengineered pinning.
\newblock {\em Phys. Rev. B}, 105(22):224409, 2022.

\bibitem{souza_soliton_2023}
J.~C.~Bellizotti Souza, N.~P. Vizarim, C.~J.~O. Reichhardt, C.~Reichhardt, and
  P.~A. Venegas.
\newblock Soliton motion induced along ferromagnetic skyrmion chains in chiral
  thin nanotracks.
\newblock {\em J. Mag. Mag. Mater.}, 587:171280, 2023.

\bibitem{du_steady_2024}
D.~Song, W.~Wang, S.~Zhang, Y.~Liu, N.~Wang, F.~Zheng, M.~Tian, R.~E.
  Dunin-Borkowski, J.~Zang, and H.~Du.
\newblock Steady motion of 80-nm-size skyrmions in a 100-nm-wide track.
\newblock {\em Nature Commun.}, 15:5614, 2024.

\bibitem{liu_strain-induced_2024}
C.~Liu, J.~Wang, W.~He, C.~Zhang, S.~Zhang, S.~Yuan, Z.~Hou, M.~Qin, Y.~Xu,
  X.~Gao, Y.~Peng, K.~Liu, Z.~Q. Qiu, J.-M. Liu, and X.~Zhang.
\newblock Strain-induced reversible motion of skyrmions at room temperature.
\newblock {\em ACS Nano}, 18(1):761--769, 2024.

\bibitem{xing_skyrmion_2022}
X.~Xing and Y.~Zhou.
\newblock Skyrmion motion and partitioning of domain wall velocity driven by
  repulsive interactions.
\newblock {\em Commun. Phys.}, 5(1):1--11, 2022.

\bibitem{souza_skyrmion_2024}
J.~C.~Bellizotti Souza, N.~P. Vizarim, C.~J.~O. Reichhardt, P.~A. Venegas, and
  C.~Reichhardt.
\newblock Skyrmion {Molecular} {Crystals} and {Superlattices} on {Triangular}
  {Substrates}, 2024.
\newblock arXiv:2407.20188 [cond-mat].

\bibitem{berdiyorov_vortex_2006}
G.~R. Berdiyorov, M.~V. Milo\v{s}evi\'{c}, and F.~M. Peeters.
\newblock Vortex configurations and critical parameters in superconducting thin
  films containing antidot arrays: Nonlinear {Ginzburg-Landau} theory.
\newblock {\em Phys. Rev. B}, 74:174512, 2006.

\bibitem{neal_competing_2007}
J.~S. Neal, M.~V. Milo\v{s}evi\'{c}, S.~J. Bending, A.~Potenza,
  L.~San~Emeterio, and C.~H. Marrows.
\newblock Competing symmetries and broken bonds in superconducting
  vortex-antivortex molecular crystals.
\newblock {\em Phys. Rev. Lett.}, 99:127001, 2007.

\bibitem{reichhardt_vortex_2007}
C.~Reichhardt and C.~J.~Olson Reichhardt.
\newblock Vortex molecular crystal and vortex plastic crystal states in
  honeycomb and kagom{\' e} pinning arrays.
\newblock {\em Phys. Rev. B}, 76(6):064523, 2007.

\bibitem{reichhardt_novel_2002}
C.~Reichhardt and C.~J. Olson.
\newblock Novel colloidal crystalline states on two-dimensional periodic
  substrates.
\newblock {\em Phys. Rev. Lett.}, 88(24):248301, 2002.

\bibitem{brunner_phase_2002}
M.~Brunner and C.~Bechinger.
\newblock Phase behavior of colloidal molecular crystals on triangular light
  lattices.
\newblock {\em Phys. Rev. Lett.}, 88(24):248302, May 2002.
\newblock Publisher: American Physical Society.

\bibitem{agra_theory_2004}
R.~Agra, F.~van Wijland, and E.~Trizac.
\newblock Theory of orientational ordering in colloidal molecular crystals.
\newblock {\em Phys. Rev. Lett.}, 93(1):018304, 2004.

\bibitem{frey_melting_2005}
A.~\ifmmode~\check{S}\else \v{S}\fi{}arlah, T.~Franosch, and E.~Frey.
\newblock Melting of colloidal molecular crystals on triangular lattices.
\newblock {\em Phys. Rev. Lett.}, 95:088302, 2005.

\bibitem{thomas_spin_2007}
A.~\ifmmode~\check{S}\else \v{S}\fi{}arlah, E.~Frey, and T.~Franosch.
\newblock Spin models for orientational ordering of colloidal molecular
  crystals.
\newblock {\em Phys. Rev. E}, 75:021402, 2007.

\bibitem{Reddy23}
A.~P. Reddy, T.~Devakul, and L.~Fu.
\newblock Artificial atoms, {Wigner} molecules, and an emergent kagome lattice
  in semiconductor moir{\' e} superlattices.
\newblock {\em Phys. Rev. Lett.}, 131(24):246501, 2023.

\bibitem{Li24}
H.~Li, Z.~Xiang, A.~P. Reddy, T.~Devakul, R.~Sailus, R.~Banerjee, T.~Taniguchi,
  K.~Watanabe, S.~Tongay, A.~Zettl, L.~Fu, M.~F. Crommie, and F.~Wang.
\newblock Wigner molecular crystals from multielectron moir{\' e} artificial
  atoms.
\newblock {\em Science}, 385(6704):86--91, 2024.

\bibitem{evans_atomistic_2018}
R.~F.~L. Evans.
\newblock Atomistic {Spin} {Dynamics}.
\newblock In W.~Andreoni and S.~Yip, editors, {\em Handbook of {Materials}
  {Modeling}: {Applications}: {Current} and {Emerging} {Materials}}, pages
  1--23. Springer International Publishing, 2018.

\bibitem{iwasaki_current-induced_2013}
J.~Iwasaki, M.~Mochizuki, and N.~Nagaosa.
\newblock Current-induced skyrmion dynamics in constricted geometries.
\newblock {\em Nature Nanotechnol.}, 8(10):742--747, October 2013.

\bibitem{paul_role_2020}
S.~Paul, S.~Haldar, S.~von Malottki, and S.~Heinze.
\newblock Role of higher-order exchange interactions for skyrmion stability.
\newblock {\em Nature Commun.}, 11(1):4756, 2020.

\bibitem{seki_skyrmions_2016}
S.~Seki and M.~Mochizuki.
\newblock {\em Skyrmions in {Magnetic} {Materials}}.
\newblock Springer International Publishing, 2016.

\bibitem{gilbert_phenomenological_2004}
T.~L. Gilbert.
\newblock A phenomenological theory of damping in ferromagnetic materials.
\newblock {\em IEEE Trans. Mag.}, 40(6):3443--3449, 2004.

\bibitem{zang_dynamics_2011}
J.~Zang, M.~Mostovoy, J.~H. Han, and N.~Nagaosa.
\newblock Dynamics of {Skyrmion} {Crystals} in {Metallic} {Thin} {Films}.
\newblock {\em Phys. Rev. Lett.}, 107(13):136804, September 2011.

\bibitem{Suppl}
See Supplemental~Material for~supplementary videos.

\end{thebibliography}

\end{document}